\documentclass[%
 reprint,
superscriptaddress,
amsmath,amssymb,
 prl,
]{revtex4-2}
\usepackage{centernot}
\usepackage{graphicx}
\usepackage{dcolumn}
\usepackage{bm}
\usepackage[dvipsnames]{xcolor}

\begin{document}


\title{Irreversibility in Non-reciprocal Chaotic Systems}

\author{Tuan Pham}
\author{Albert Alonso}
\author{Karel Proesmans}
\affiliation{Niels Bohr Institute, University of Copenhagen,
Blegdamsvej 17, Copenhagen, 2100-DK, Denmark
}


\begin{abstract}
 How is the irreversibility of a high-dimensional chaotic system controlled by the heterogeneity in the non-reciprocal interactions among its elements? In this paper, we address this question using a stochastic model of \emph{random}
 recurrent 
 neural networks that undergoes a transition from quiescence to chaos at a critical heterogeneity. In the thermodynamic limit, using dynamical mean field theory,  we obtain an exact expression for the averaged entropy production rate  -- a measure of irreversibility -- for any heterogeneity level $J$. We show  how this quantity becomes a constant at the onset of chaos, while changing  its functional form upon crossing this point. The latter can be elucidated by closed-form approximations valid for  below and slightly above the critical point and for large $J$. 
\end{abstract}

\maketitle

\emph{Introduction.} Complex systems of heterogeneously interacting agents are 
intrinsically out-of-equilibrium. Underlying their emergent behaviours is the intricate interplay between stochasticity, non-linearity and nonreciprocity of interactions. To  understand the dynamics of these systems, a general framework known as dynamical mean field theory (DMFT) \cite{Martin, Dominicis,COOLEN2001, Chow_2015, Hertz2017,Marti2018, Cugliandolo2024} has been developed. Significant progress has been made in a variety of problems through DMFT, including the relaxation of $p$-spin glasses towards  weak ergodicity-breaking states \cite{Cugliandolo_Kurchan},  the dynamics of random replicator \cite{Opper1992} and the effective noise of stochastic gradient descent \cite{Mignacco_2022}, to name but a few. A renewed interest in applying DMFT to complex systems, such as neural networks \cite{clark2023theory, Keup, Schuecker2018,Mastrogiuseppe2017,martorell2023dynamically}, ecology \cite{Bunin,  Roy2019, depirey2024criticalbehaviorphasetransition, galla2024generatingfunctionalanalysisrandomlotkavolterra}, biological adaptation \cite{pham2024dynamical}, oscillatory dynamics \cite{shmakov2023coalescence, pruser2023nature}, and active matter \cite{silvano2024emergent, Paoluzzi2020}, has emerged recently. 

Meanwhile, 
the thermodynamics of non-equilibrium phase transitions has been studied in a wide variety of systems, including active matters \cite{Agranov_2022,Ferretti, Yu2022, GrandPre, Alston}, spin models with order-disorder transitions \cite{Tome2012, Noa2019, Martynec_2020, Tome2023, OLIVEIRA2024}, other systems with synchronisation transitions \cite{Zhang2020}, Hopf bifurcations \cite{fritz2020stochastic}, the onset of limit cycles \cite{HerpichPRX} and pattern formations \cite{falasco2018information,rana2020precision}. Meanwhile, progress on incorporating a thermodynamic treatment into DMFT has so far been limited to discrete-time systems \cite{Aguilera2023}. 
A proper understanding of  the onset of chaos -- one of the most important classes of non-equilibrium phase-transitions, is ,  to the best of our knowledge, currently lacking.

The goal of this letter is  two-fold. On the one hand, we want to integrate stochastic thermodynamics into the DMFT treament of  continuous-time dynamics. This will not only deepen our understanding of the energetic cost associated with operating complex and disordered systems, but also facilitate the  applications of general results from stochastic thermodynamics, such as the thermodynamic uncertainty relations \cite{barato2015thermodynamic,horowitz2020thermodynamic} and speed-limits \cite{aurell2011optimal,van2023thermodynamic}, to those systems. On the other hand,  we 
aim to build a thermodynamic theory of chaotic systems that can quantify their behaviours using entropy production.

We illustrate these ideas using an important  model of random neural networks  -- sometimes referred to as the ``Wilson–Cowan'' model  \cite{Buice_2013}. This model  has been widely used to study machine-learning algorithms \cite{Jaeger2004} and, by means of DMFT \cite{Sompolinsky,   Crisanti2018, Cabana, maclaurin2024population},  a transition from fixed-point regime to chaos beyond a critical heterogeneity level of synaptic weights has been established. Such a transition is believed to be particularly relevant for understanding the computational capabilities of neural networks with an optimal information processing capacity at the edge of chaos \cite{Toyoizumi2011,Sussillo2009} and the cost of this information processing itself \cite{Alessandro}. Our work can shed light on the question of how the energy dissipation, as quantified by  the  entropy production rate, changes at this transition.
Though we shall focus on a thermodynamically-consistent model of  neural networks, we  need to stress that our methodology can  be extended to other systems amenable to DMFT analysis, such as active matter and ecological models.

\emph{Model.} We consider a system of $N$ neurons,   each characterized by a continuous variable $x_i(t)$ and has a saturating nonlinear input-output transfer function $f$. We choose $f(\cdot)={\rm tanh} (\cdot)$. 
They are
coupled through a fully-connected network of interactions $\mathbf{J}$ without self-loop (i.e. $J_{ii}=0$, $\forall\, i$).
 The  dynamics of these units are modelled as Langevin equations,
\begin{equation}
    \frac{dx_i}{dt} = -x_i + {\rm tanh}\Big(\sum_{j=1, j\neq i}^N J_{ij} x_j \Big) +  \xi_i(t)
    \label{model}
\end{equation}
where $\xi_i(t)$ is a zero-mean white noise  with corelator $C_{\xi}(t,t')\equiv\langle \xi_i(t)\xi_j(t') \rangle_{\xi} = \sigma^2\delta_{ij}\delta(t-t')$, which corresponds to a thermal noise from the environment and is related to the environment's temperature $T$ as $k_BT=\sigma^2/2$, with $k_B$  the Boltzman constant. 
Here we consider a \emph{quenched} ensemble of network configurations with asymmetric (i.e. uncorrelated) independent Gaussian couplings $J_{ij}$, whose mean, variance and covariance are
\begin{equation}
     \big[ J_{ij} \big]_{\mathbf{J}} =\frac{\mu}{N}\,;\Big[  J_{ij}^2\Big]_{\mathbf{J}}  = \frac{J^2}{N}\,; \Big[ \Big(J_{ij} -\frac{\mu}{N}\Big)\Big(  J_{ji} - \frac{\mu}{N}\Big)\Big]_{\mathbf{J}}  = 0   \label{mean_variance}
\end{equation}
 Hereafter  $[\cdot]_{\mathbf{J}}$, referred to as ensemble average, denotes averaging with respect to the coupling distribution $P(\mathbf{J})= \prod_{i\neq j}P(J_{ij})$. 
 In this paper we will consider only the case of $\mu=0$. The network is then called `balanced', as  each neuron, on average, has a net zero input due to a balance between excitation and
inhibition. Though we use  Gaussian distributions for the coupling strengths, distributions that include a $\delta$ function at zero, corresponding to sparse connections, or that have discrete support, corresponding to a finite number of possible  strengths, give the same results in the  large $N$ limit \cite{Schuecker2018}. 

We note that Eq.~\eqref{model} does not satisfy detailed balance due to the non-reciprocity ($J_{ij}\neq J_{ji}$) and due to the non-linearity of the input-output transfer function. The system is hence out of equilibrium and constantly produces entropy, even in steady state. 
In the low-noise limit, $\sigma\rightarrow 0$, it exhibits two different phases, namely a disordered phase, where the dynamics evolves to a fixed point, for $J<1$ and a chaotic phase, for $J>1$,  where all neurons keep fluctuating despite the absence of noise.

\emph{Entropy production rate.} We want to compute the entropy production rate of the system described in Eqs.~\eqref{model}-\eqref{mean_variance}, using the following  decomposition of  the total entropy production rate according to  stochastic thermodynamics \cite{seifert2012stochastic}:
\begin{equation*}
   \dot{S}_{\rm tot}= \dot{S}_{\rm res}+\dot{S}_{\rm sys}
\end{equation*}
with
\begin{eqnarray}
     \dot{S}_{\rm res}&=\displaystyle \frac{2}{\sigma^2}\sum_{i} F_i(\mathbf{x})\circ \frac{dx_i}{dt}\,,\,\, F_i(\mathbf{x})\equiv -x_i + {\rm tanh}\big([\mathbf{Jx}]_i \big)\,\,\,\label{entropy_reservoir}\\
        \dot{S}_{\rm sys}&= \displaystyle -k_B\sum_i\left(\frac{\partial}{\partial x_i} \ln p(\mathbf{x},t)\right)\circ \frac{d x_i}{dt}-k_B\frac{\partial}{\partial t}\ln p(\mathbf{x},t)\,\,
\end{eqnarray}
the rate at which  entropy is dissipated into the environment and the change in systems entropy, respectively. Here ``$\circ$" denotes the Stratonovich convention and $p(\mathbf{x},t)$ is the  joint probability distribution of all neurons at time $t$.
At steady state, the average system entropy production rate vanishes, $\langle \dot{S}_{\rm sys}^{\rm (ss)} \rangle = 0$, and hence 
$\langle \dot{S}_{\rm tot}^{\rm (ss)}\rangle =\langle \dot{S}_{\rm res}^{\rm (ss)}\rangle$, where the $\langle\cdot\rangle$ represent the average taken over all noise-realisations. Therefore, we shall  focus on calculating the entropy production in the reservoir, $ \dot{S}_{\rm res}$ for the remainder of this paper.

\emph{DMFT description.}
In the $N\rightarrow \infty$ limit, the system is self-averaging, i.e., its macroscopic behaviour can be assumed to depend only on the statistical properties of the disorder in Eq.~\eqref{mean_variance}. 
Using DMFT, we obtain one-dimensional effective dynamics that become an \emph{exact} description of the  dynamics Eq. \eqref{model} as $N\rightarrow \infty$:
\begin{equation}
 \dot{x} = -x(t) + u(t) +\xi(t) \,,\qquad u \equiv {\rm tanh}\big(  \eta(t) \big)\label{effective_dynamics}
\end{equation}
where we have introduced $\eta$ -- a Gaussian noise with zero mean and whose correlation $C_\eta(t,t')\equiv\langle \eta(t)\eta(t')\rangle$,  needs to be defined self-consistently as
\begin{equation}    C_\eta(t,t')= J^2 C_x(t,t')    \label{eta_correlator}
\end{equation}
with
\begin{equation}
   C_x(t,t') \equiv \big\langle x(t)x(t') \big \rangle \label{x_correlator}
\end{equation}
One can interpret $x(t)$ as the representative neuron of a homogeneous population, in which all neurons with statistically identical connections and dynamical properties become identical after taking the ensemble average. Therefore, we have \begin{equation}
    \left[\left\langle x_k(t)\right\rangle\right]_{\mathbf{J}}=\left\langle x(t)\right\rangle^{(MF)},\qquad \forall k=1,..,N.
\end{equation}
The superscript $(MF)$ denotes the average taken over the mean-field dynamics  Eq. \eqref{effective_dynamics}. Though  Eq.~\eqref{effective_dynamics} is a linear equation in terms of $x(t)$, the non-linearity of the full system Eq.~\eqref{model} is encoded in the  self-consistency relation Eq.~\eqref{eta_correlator} [note the difference with the celebrated model in \cite{Sompolinsky} whose effective dynamics are given by $(\partial_x+1)x(t)= \eta(t)+\xi(t)$  with  $C_{\eta}(t,t')= J^2 \langle {\rm tanh}(x(t)){\rm tanh}(x(t'))\rangle$]. 

In appendix A, we show that this equivalence  between the exact and the DMFT dynamics can be extended to the entropy production into the environment,
\begin{eqnarray}
\left[\left\langle\dot{s}_{\rm res}(t)\right\rangle \right]_{\mathbf{J}}&=&\Big\langle\dot{s}^{(MF)}_{\rm res}(t)\Big\rangle,
    \label{eq:SEQ}
\end{eqnarray}
with $\dot{s}_{\rm res}(t)=\dot{S}_{\rm res}(t)/N$ and
\begin{equation}
\dot{s}^{(MF)}_{\rm res}(t)=(-x(t)+u(t))\circ \frac{dx(t)}{dt}.
\label{numerical_simulation_DMFT_entropy}
\end{equation}
We dropped the superscript over the average for notational simplicity. This allows us to
 conclude that the average entropy flux into the environment per neuron in the original system is exactly the same as that of the DMFT one. Hereafter, we will focus 
 on calculating the average entropy production rate through this DMFT approach. 
 
\begin{figure}
\includegraphics[scale=0.8]{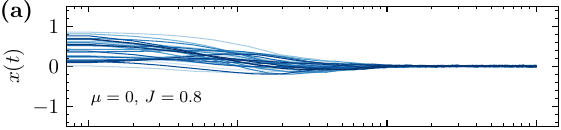} 
\includegraphics[scale=0.8]{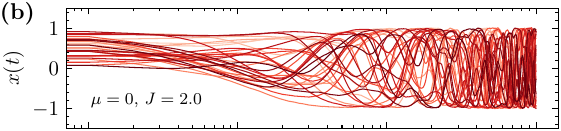} 
\includegraphics[scale=0.8]{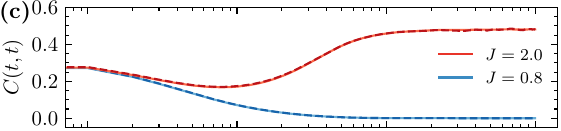} 
\includegraphics[scale=0.8]{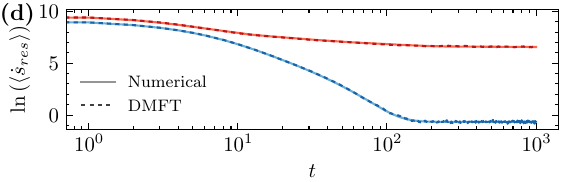} 
   \caption{\textbf{(a)-(b)}  Time series of neuronal states evolving according to Eq.~\eqref{model} in  the trivial-fixed-point regime ($J=0.8$); 
   and the chaotic regime ($J=2$). \textbf{(c)-(d)} Equal-time correlation $C(t,t)$ and  the entropy production rate for   the trivial-fixed-point regime (blue) 
   and in the chaotic regime (red). $N=1000$ neurons for the numerics and $N_{\rm traj}=1000$ trajectories are used for the numerical integration of DMFT. 
   In all panels, $\sigma=0.01$ and $x(0)$ is uniformly distributed in $[0,1]$. }
    \label{fig:fig1}
\end{figure}

 Before continuing a (semi-)analytical calculation of the entropy production rate, we numerically verify the agreement between simulations of the  dynamics Eq.~\eqref{model} and  its DMFT description given in Eq.~\eqref{effective_dynamics}. In Fig.~\ref{fig:fig1} \textbf{(c)}-\textbf{(d)}  an excellent agreement between the two dynamics is found  for both $C_x(t,t)$ and $\left\langle \dot{s}_{\rm res}(t)\right\rangle$ even before reaching  steady states both for the fixed-point dynamics (Fig.~\ref{fig:fig1} \textbf{(a)}) as for the chaotic phase (Fig.~\ref{fig:fig1} \textbf{(b)}).

In appendix B, 
we show that under the steady-state conditions, for which $C_x(\tau)= \lim_{t'\rightarrow \infty}C_x(t',t'+\tau)$, the entropy production rate simplifies to
\begin{equation}
    \big\langle  \dot{s}^{\rm (MF)}_{\rm res}\big\rangle  = -\frac{2}{\sigma^2} \ddot{C}_{x}(0^+)+1,\label{eq:dsc}
\end{equation}
which means that knowledge of the auto-correlation function of a single neuron $x$ obeying Eq. \eqref{effective_dynamics} is sufficient to calculate the entropy production rate of the system. According to \cite{Schuecker2018}, the curvature $\ddot{C}_{x}(0^+)$ changes its sign from positive to negative once crossing the \emph{true} critical point    $J^{*}$, resulting in $\big\langle  \dot{s}^{\rm (MF)}_{\rm res}\big\rangle<1$  and $\big\langle  \dot{s}^{\rm (MF)}_{\rm res}\big\rangle>1$, below and above $J^{*}$, respectively. At $J=J^{*}$,  
$\ddot{C}_x(0^+)=0$,  yielding
\begin{equation}
  \big\langle  \dot{s}^{\rm (MF)}_{\rm res}\big\rangle_{J=J^{*}}  =1    \label{entropy_critical}
\end{equation}

\begin{figure}[t]
\includegraphics[scale=0.7]{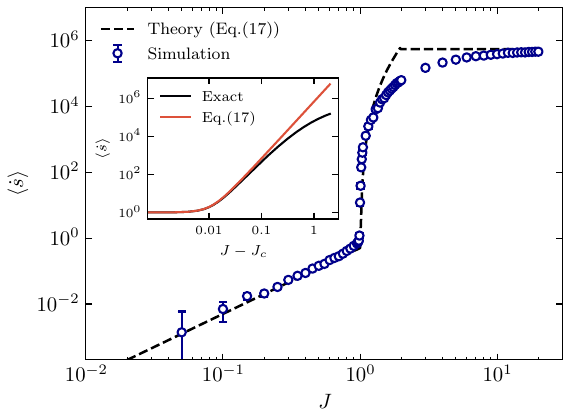} 
\caption{Comparison of $\langle \dot{s}_{\rm res}^{\rm (ss)}\rangle$ from simulations with $\sigma=0.001$, $N=1000$ and the asymptotics given by Eq.~\eqref{eq:appr}. The inset shows the comparison between the  entropy production computed by the exact numerics of DMFT and the asymptotics near $J\approx J_c = 1+ 3^{1/4}\sigma$.}
\label{fig:fig2}
\end{figure}

\emph{Solution of the DMFT equation.} Following the pioneering approach in  \cite{Sompolinsky, Crisanti2018}, we map the dynamics of $C_x(\tau)$  onto that of a \emph{classical} particle moving under a potential $V$ parametrized by the initial condition $C_0\equiv C_x(0)$. As detailed in appendix C,  for $\tau\geq 0$, one can write (cf.~Appendix C):
\begin{equation}\begin{aligned}\partial^2_\tau C_x(\tau) &= - \frac{\partial}{\partial C_x(\tau)} \,V(C_x(\tau)|C_0) - \sigma^2 \delta(\tau)  \\ V(C_x(\tau)|C_0) &= \frac{\big\langle \Omega(\eta(0))\Omega(\eta(\tau))\big\rangle_\eta  -\big\langle \Omega(\eta(0))\big\rangle^2_\eta}{J^2} -\frac{C_x(\tau)^2}{2} 
\label{EOM_correlation} 
\end{aligned}
\end{equation}
where $\Omega(z) := \ln{\rm cosh}(z)  = \int_0^z  {\rm tanh}(y) dy$ and
 \begin{multline}
     \big\langle \Omega(\eta(0))\Omega(\eta(\tau))\big\rangle_\eta\\=\int \int d\eta_1 d\eta_2\,\frac{e^{-\frac{\left(C_0(\eta_1^2+\eta_2^2)-2C_x(\tau)\eta_1\eta_2\right)}{2J^2(C_0^2-C_x(\tau)^2)}}}{2\pi J^2\sqrt{C_0^2-C_x(\tau)^2}}\,\Omega(\eta_1)\Omega(\eta_2),\label{eq:omav}
 \end{multline}
 and a similar equation for $\big\langle \Omega(\eta(0))\big\rangle_\eta$. One can find $C_0$ through the following equations: 
\begin{equation}
V(C_0|C_0) = -\frac{\sigma^4}{8},\qquad \dot{C}_x(0^+)=-\frac{\sigma^2}{2} .
\label{initial_condition}
\end{equation}

The entropy production rate given in Eq.~\eqref{eq:dsc} can also be rewritten in terms of the potential $V$ as
\begin{equation}
    \left\langle \dot{s}^{(MF)}_{\rm tot}\right\rangle=\frac{2}{\sigma^2} \left.\frac{\partial}{\partial C_x(\tau)} \,V(C_x(\tau)|C_0)\right|_{C_x(\tau)=C_0}+1.
    \label{eq:dsV}
\end{equation}
This equation provides a  numerical procedure to calculate  exactly the entropy production rate of the network in the thermodynamic limit: we use the first one in Eq.~\eqref{initial_condition} to find $C_0$, then apply Price's theorem to calculate $ V(C_x|C_0)$ by Eq.~\eqref{EOM_correlation} and finally compute the entropy production rate  by Eq.~\eqref{eq:dsV}. Hereafer we shall refer this procedure as \emph{the exact numerics of DMFT} to distinguish it from the other scheme consisting of numerically integrating the effective dynamics Eq.~\eqref{effective_dynamics} and computing the associated $\left\langle \dot{s}^{(MF)}_{\rm tot}\right\rangle$ by simulating Eq.~\eqref{numerical_simulation_DMFT_entropy}. 

\emph{Limits and asymptotics}  To get more intuition about the entropy production rate, we shall derive a number of analytical results from  solutions to the above set of equations in some limits. In appendix D, we show that
\begin{widetext}\begin{equation}\big\langle  \dot{s}^{(MF)}_{\rm tot}\big\rangle\simeq
\begin{cases}
    \frac{J^2}{2} & J\ll1 
    \\
    \frac{2}{3\sigma^2} \left[\frac{  (J-1)^2}{2} + \frac{1}{2}\,\sqrt{(J-1)^4 -3\sigma^4 }\right]^{1/2}\sqrt{(J-1)^4 -3\sigma^4} +1 & 0\leq J-J_c\ll 1,\textrm{ and } \sigma^2\ll 1\\
    \frac{4}{\pi\sigma^2}\left(1-\sqrt{\left(1-\frac{\pi}{2}\right)^2+\left(\frac{\pi\sigma^2}{4}\right)^2}\right)+1 & J\gg 1
\end{cases},\label{eq:appr}
\end{equation}
\end{widetext}
where in the small-noise limit $J_c=1+3^{\frac{1}{4}}\sigma$. $J_c$ is  the lowest order approximation of the true critical point $J^{*}$. 

We compare our analytical approximation in Eqs.~\eqref{eq:appr}
with direct numerical simulations of Eq.~\eqref{entropy_reservoir} at a fixed noise strength $\sigma=0.001$ in Fig.~\ref{fig:fig2}. We verify that $\big\langle  \dot{s}_{\rm tot}\big\rangle\approx J^2/2$ up to values relatively close to $J=1$. Subsequently, the second and third lines of Eq.~\eqref{eq:appr}  capture the  transition into the chaotic regime. We also find that entropy production rate does indeed level of at the value predicted by the last line of Eq.~\eqref{eq:appr} around $J\approx 10$.
The inset shows the comparison between our approximation Eq.~\eqref{eq:appr}  of the entropy production rate and its exact calculations obtained from Eqs.~\eqref{EOM_correlation}-\eqref{eq:dsV} for $J$ near $J_c$. One can see that our approximation, Eq.~\eqref{eq:appr}, is indeed in good agreement with the exact dynamics and that the entropy production goes to $1$ as $J$ approaches $J_c$.

\begin{figure}
\includegraphics[scale=0.7]{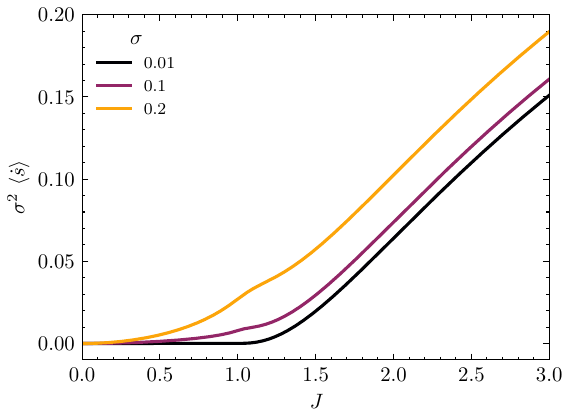} 
\caption{$\sigma^2\left\langle \dot{s}^{(MF)}_{\rm tot}\right\rangle$ as a function for $J$ for various choices of $\sigma$. One can observe a clear change in the behavior of this quantity around $J\approx 1$, signalling a phase-transition. Here $\Big\langle \dot{s}^{(MF)}_{\rm tot}\Big\rangle$ is computed using the exact numerics of DMFT.}  \label{fig:fig3a}
\end{figure}
As mentioned in the beginning of this paper, this model exhibits two phases in the low-noise limit, $\sigma\rightarrow 0$: a disordered phase for $J<1$ and a chaotic phase for $J>1$. From Eq.~\eqref{eq:appr} we can conclude that these two regimes correspond to qualitatively very different entropy production rates $\big\langle  \dot{s}_{\rm tot}\big\rangle$: in the disordered phase, $\big\langle  \dot{s}_{\rm tot}\big\rangle$ essentially settles on a finite value independent of the thermal noise strength; meanwhile, in the chaotic phase, $\big\langle  \dot{s}_{\rm tot}\big\rangle$ diverges as the noise becomes negligible. It is then  natural  to ask whether this transition is a genuine thermodynamic phase-transition. The framework of stochastic thermodynamics suggests that first-order phase-transition generally have a discontinuity in $\sigma^2\left\langle \dot{s}_{\rm tot}\right\rangle$, while higher order phase-transitions have a discontinuity in their derivatives with respect to $J$ \cite{nguyen2020exponential}. In Fig.~\ref{fig:fig3a}, we observe that $\sigma^2\left\langle \dot{s}_{\rm tot}\right\rangle$ vanishes as $\sigma\rightarrow 0$ for $J<J_c$, whereas it continuously changes to a finite value for $J> J_c$. This indeed suggests a  continuous transition, in agreement with the literature showing a smooth increase from zero of the largest  Lyapunov exponent \cite{Engelken2023} and the critical slowing down \cite{Schuecker2018}   at the onset of chaos.

\begin{figure}
\includegraphics[scale=0.7]{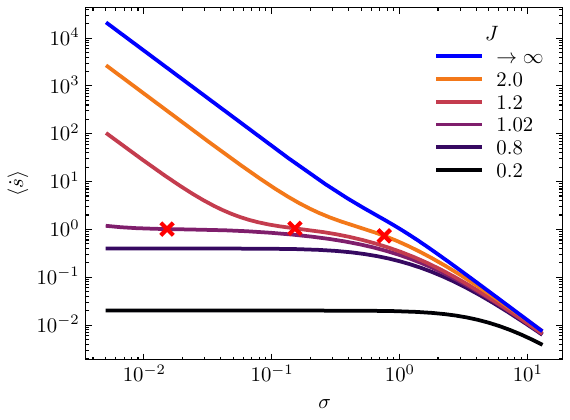}  
\caption{Entropy production rate as function of $\sigma$ below  the transition to chaos $J=0.2, 0.8$ and it $J=1.02,1.2, 2$ and $J\rightarrow\infty$ using the exact numerics of DMFT.
The \emph{approximate} critical point $\sigma=3^{-\frac{1}{4}}(J-1)$ are denoted with red crosses.}
    \label{fig:fig3}
\end{figure}

In Fig.~\ref{fig:fig3}, we show that the signatures of this phase-transition carry over to finite noise levels. One can verify that for $J<1$ levels off at relatively high noise-levels ($\sigma\sim 10^{-1}$) and does no longer increase when $\sigma$ decreases. For $J>1$ on the other hand, one can see that there is an intermediate levelling off, but that the entropy production rate starts increasing again for $\sigma<3^{-1/4}(J-1)$, denoted  by red crosses, corresponding to $J=J_c$. If one decreases the noise beyond this point, one observes that the entropy production rate starts scaling as $\sigma^{-2}$ again.

\emph{Conclusions.}
In this paper, we study the entropy production rate in a random recurrent neural network that undergoes a transition from fixed-point to chaotic regime with increasing coupling heterogeneity. Crucially, our analysis is based on applying  DMFT to the study of the system's entropy production rate. It is possible  to apply this approach to more complex dynamics with multiple fixed points \cite{Breffle2023}, suppression of chaos \cite{Rajan,Engelken2022, Takasu}, structural chaos \cite{Mastrogiuseppe2018} and non-monotonic dependence on noise \cite{Garnier}, or to study the entropy production rate in other types of networks such as sparse directed networks \cite{metz2024dynamical},  deep neural networks  \cite{Bahri, Segadlo_2022}, 
 networks with  heterogeneous degree distribution   \cite{aguirre, park2024incorporating, poley2024interaction},  multiple populations \cite{Kadmon2015, Aljadeff,
Landau2021}, low-rank structures \cite{Schuessler}, or  synaptic plasticity \cite{clark2023theory}. In particular, this might lead to design principles for minimizing the thermodynamic cost of operating a diverse set of neural networks.

Another interesting research direction would be to check whether the results from this letter can be extended to other chaotic systemms, such as logistic maps \cite{baldovin2005parallels}, or active nematics \cite{venkatesh2022distinct}. In particular, it would be interesting to check whether one can find similar scaling of entropy production rate with noise in the chaotic phase. If so, this could lead to universal thermodynamic bounds and constraints at the onset of chaos -- a first crucial step towards a general framework for quantifying non-equilibirum phase transion in chaotic systems.

Finally, of paramount importance is the question of finding an optimal protocol that minimizes the  amount of heat necessary to drive a system towards its chaotic regime. This would require an extension of   our analysis to systems with time-dependent driving. For this kind of problems, one could use the recently developed linear-response framework \cite{sivak2012thermodynamic,davis2024active}.

\begin{acknowledgments}
Albert Alonso acknowledges support from the Novo Nordisk Foundation  Grant Agreement NNF20OC0062047. Karel Proesmans is funded by the European Union’s Horizon 2020 research and innovation program under the Marie Sk\l odowska-Curie grant agreement No. 101064626 `TSBC’, and from the Novo Nordisk Foundation (grant No. NNF18SA0035142 and NNF21OC0071284).
We  would like to thank Carlos E. Fiore, Etienne Fodor, Samantha Fournier, Prashant Singh and Jonas Berx  for stimulating discussions.
\end{acknowledgments}

\bibliography{Doublereplica}

\newpage

\onecolumngrid




\newpage

\section{Appendix A: Derivation of Eq. (5) and Eq. (10)}
Following the standard presentation in \cite{COOLEN2001, Buice_2013,Chow_2015, Hertz2017,Marti2018}, we derive DMFT for a \emph{quenched} ensemble of random matrices
in Eq.~\eqref{mean_variance}.
 Let   $\langle \cdot \rangle$ in this section denote the average taken with respect to the measure $ \mathbb{P}\big(\{\bold{x}\}_{0,t_f}\big)$ -- the distribution  of an ensemble of trajectories $\{\bold{x}\}_{0,t_f}$, from $\bold{x}_0=\bold{x}(0)$ at  $t =0$ to $\bold{x}_f = \bold{x}(t_f)$ at $t = t_f$. We shall assume that the system is in steady state and therefore take  $P (\mathbf{x}(0))$ as the steady-state distribution. 

 The probability to observe a trajectory $\{\mathbf{x}\}_{0,t_f}$ is then given by 
\begin{equation}
    \mathbb{P}\big(\{\bold{x}\}_{0,t_f}\big)=\int D\big[\hat{x}f\hat{f}\big] \exp\left\{i \sum_{k=1}^N \int dt \Big[S^{(0)}_k -   \sum_{j=1}^N J_{kj} \hat{f}_k x_j\Big]\right\} 
\end{equation}
 where $D[\hat{x}f\hat{f}] :=  \prod_{t=0}^{t_f}\prod_{k=1}^{N}   D\hat{x}_k(t) Df_k(t)  D\hat{f}_k(t)$ is  the functional measure over all possible paths, and where
 \begin{equation}  S^{(0)}_k(t) =\hat{x}_k \Big[(\partial_t +1)x_k(t)  - {\rm tanh}\big(f_k(t)  \big) \Big]+\, i\sigma^2 \hat{x}^2_k(t)/2  + \hat{f}_k(t) f_k(t) 
\label{free_action}
\end{equation}
is the free Lagrangian for a neuron $k$. 
Using the definition, Eq.~\eqref{eq:SEQ}, one can immediately see that
\begin{equation}
    Z_S\left[\mathbf{\lambda}\right]=\left[ \int D\big[x\hat{x}f\hat{f} \big] \exp\left\{\Omega - i\sum_{j \neq{k}}^N J_{kj} \int dt    \hat{f}_k x_j\right\}\right]_{\mathbf{J}}\label{generating_function_for_the_entropy_2} 
\end{equation}
with
\begin{equation}
    \Omega = i\sum_{k=1}^N \int dt \Big[ S^{(0)}_k + \lambda(t)  \dot{x}_k(t)\big[{\rm tanh}\big(f_k(t)\big) - x_k(t)\big] 
    \Big].
\end{equation}
To make further progress, we will now explicitly take the average over $\mathbf{J}$. Firstly, we note that only the last term in the exponential of Eq.~\eqref{generating_function_for_the_entropy_2} depends on $\mathbf{J}$ and that the distribution of $\mathbf{J}$ is Gaussian with covariances given by Eq.~\eqref{mean_variance}. This can be used to get
\begin{eqnarray}
    Z_S\left[\mathbf{\lambda}\right]&=&\int D\big[x\hat{x}f\hat{f} \big] \exp\left\{\Omega\right\} \left[\exp\left\{ - i\sum_{j \neq{k}}^N J_{kj} \int dt    \hat{f}_k x_j\right\}\right]_{\mathbf{J}}\nonumber\\
    &=&\int D\big[x\hat{x}f\hat{f} \big] \exp\left\{\Omega-\frac{J^2}{2N}\int dt dt' \sum_{j\neq k}^NA_{kj}(t,t')\right\}
\end{eqnarray}
where
\begin{equation}
 A_{kj}(t,t') =\hat{f}_k(t) \hat{f}_k(t')  x_j(t)x_j(t') 
\end{equation}

We now introduce new observables:
\begin{subequations}
\label{allequations7}
 \begin{eqnarray}
q(t,t') &= &  \frac{1}{N}\sum_{k} x_k(t) x_k(t') \label{auxiliary4}
\\ Q(t,t') &= & \frac{1}{N}\sum_{k} \hat{f}_k(t) \hat{f}_k(t'). \label{auxiliary5}
\end{eqnarray}
\end{subequations}
These definitions can be used to write
\begin{multline}
    Z_S\left[\mathbf{\lambda}\right]= \int D\big[x\hat{x}f\hat{f} \big]\int D\big[Qq\big] \exp\left\{\Omega-\frac{NJ^2}{2}\int dt dt' q(t,t')Q(t,t')\right\}\delta\left(Nq(t,t')-\sum_{k} x_k(t) x_k(t')\right)\\\times\delta\left(NQ(t,t')-\sum_{k} \hat{f}_k(t) \hat{f}_k(t')\right)\label{eq:Zsexact}
\end{multline}
where $D\big[Qq\big]=Dq(t,t')DQ(t,t')/N^2$
We can now use the identity
\begin{equation}
    \delta\left(Nq-\sum_kx_k(t)x_k(t')\right)=\frac{1}{2\pi}\int d\hat{q} e^{i\int dt\int dt'\hat{q}(t,t')(Nq(t,t')-\sum_kx_k(t)x_k(t'))}
\end{equation}
to write
\begin{eqnarray}
    Z_S\left[\mathbf{\lambda}\right]&=& \int D\big[Q\hat{Q}q\hat{q}\big] \exp\left\{N\left(g\left(q,\hat{q},Q,\hat{Q}\right)+\frac{1}{N}\log\int D\big[x\hat{x}f\hat{f} \big]\exp\left(\mathcal{L}\left(x,\hat{x},f,\hat{f},\hat{q},\hat{Q}\right)\right)\right)\right\},
\end{eqnarray}
with $D\big[Q\hat{Q}q\hat{q}\big]=D\big[Qq\big]D\hat{q}(t,t')D\hat{Q}(t,t')/(4\pi^2)$,
\begin{eqnarray}  g\left(q,\hat{q},Q,\hat{Q}\right)&=&i\int dt\, dt' \big[\hat{q}q + \hat{Q}Q \big]- \frac{J^2}{2} \int dt\, dt'qQ \\
    \mathcal{L}\left(x,\hat{x},f,\hat{f},\hat{q},\hat{Q}\right)&=&\Omega - i \sum_{k} \int dt dt' \big[\hat{q}(t,t') x_k(t)x_k(t')  + \hat{Q}(t,t') \hat{f}_k(t)\hat{f}_k(t')  \big] .
\end{eqnarray}
Up to this points, all calculations have been exact and we have not done any approximations. To make further progress, we will now do a mean-field approximation. In particular, we note that for large $N$ the integral in Eq.~\eqref{eq:Zsexact} is completely dominated by values for which the term inside the exponential is maximized. Therefore, we can do a sadle-point approximation \cite{Hertz2017}, to arrive at
\begin{equation}
    Z_S[\lambda]\underset{N\rightarrow \infty}{=} Z_{S}^{(MF)}[\lambda]=\max_{q,\hat{q},Q,\hat{Q}} \exp\left\{N\left(g\left(q,\hat{q},Q,\hat{Q}\right)+\frac{1}{N}\log\int D\big[x\hat{x}f\hat{f} \big]\exp\left(\mathcal{L}\left(x,\hat{x},f,\hat{f},\hat{q},\hat{Q}\right)\right)\right)\right\}
\end{equation}
One can now take the derivatives with respect to $q,\hat{q},Q$ and $\hat{Q}$ to find their optimal values. This gives
\begin{eqnarray}
    \hat{q}(t,t')&=&-\frac{iJ^2}{2}\,Q(t,t')\\
    \hat{Q}(t,t')&=&-\frac{iJ^2}{2}\,q(t,t')\\
    q(t,t') &= &  \frac{1}{N}\sum_k\left\langle x_k(t) x_k(t')\right\rangle_{\mathcal{L}}\label{eq:qres}\\ 
    Q(t,t') &= & \frac{1}{N}\sum_{k} \left\langle\hat{f}_k(t) \hat{f}_k(t')\right\rangle_{\mathcal{L}},\label{eq:Q2res}
\end{eqnarray}
where
\begin{equation}
    \left\langle X\right\rangle_{\mathcal{L}}=\frac{\int D\big[x\hat{x}f\hat{f} \big]X(x,\hat{x},f,\hat{f})\exp\left(\mathcal{L}\left(x,\hat{x},f,\hat{f},\hat{q},\hat{Q}\right)\right)}{\int D\big[x\hat{x}f\hat{f} \big]\exp\left(\mathcal{L}\left(x,\hat{x},f,\hat{f},\hat{q},\hat{Q}\right)\right)}
\end{equation}
for any observable $X$.
This means that the moment-generating function can be written as
\begin{multline}    Z_S^{(MF)}=\int D\big[x\hat{x}f\hat{f} \big]\exp\left(\frac{NJ^2}{2}\int dt dt'\,q(t,t')Q(t,t')+i\sum_k\int dt\left[S^{(0)}_k(t)+\lambda(t)\dot{x}_k(t)\Big\{{\rm tanh}\big(f_k(t)\big) - x_k(t)\Big\}\right]\right.\\\left.-\frac{J^2}{2}\int dt dt' \left[Q(t,t') x_k(t)x_k(t')  + q(t,t')\hat{f}_k(t)\hat{f}_k(t') \right]\right).
\end{multline}
One can now see that all neurons $k$ and $k'$ are independent of each other. Therefore, this can be rewritten as
\begin{multline}
Z_S^{(MF)}=\left(\int D\big[x\hat{x}f\hat{f} \big]\exp\left\{i\int dt\left[S^{(0)}(t)+\lambda(t)\dot{x}(t)\big\{{\rm tanh}\big(f(t)\big) - x(t)\big\}\right]\right.\right.\\\left.\left. +\frac{J^2}{2}\int dt dt'\left[Q(t,t')\big(q(t,t')- x(t)x(t') \big) - q(t,t')\hat{f}(t)\hat{f}(t') \right]\right\}\right)^N,
\end{multline}
where we replaced the $N$-dimensional vector $\mathbf{x}$ by a single neuron $x$ and similar for $\hat{x}$ $f$, $\hat{f}$.
Before continuing, we note that, for $\lambda=0$, $q(t,t')$ and $Q(t,t')$ have solutions given by 
\begin{equation} q(t,t')=\left\langle x(t)x(t')\right\rangle^{(MF)},\qquad Q(t,t')=\left\langle \hat{f}(t)\hat{f}(t')\right\rangle^{(MF)}.\label{eq:qsollam0}
\end{equation}
Here, the superscript $(MF)$ stands for the average taken over the probability distribution
\begin{equation}    \mathbb{P}\big(\{{x}\}_{0,t_f},\{{f}\}_{0,t_f}\big)=\int D\big[\hat{x}\hat{f} \big]\exp\left(i\int dt S^{(0)}(t) -\frac{1}{2}\int dtdt'q(t,t')\hat{f}(t)\hat{f}(t') \right)
\,.\label{eq:Pdef}
\end{equation}
One can easily verify that these relations are indeed a self-consistent solution of Eqs.~\eqref{eq:qres}-\eqref{eq:Q2res}. We can use the normalisation condition \cite{galla2024generatingfunctionalanalysisrandomlotkavolterra} to  further obtain, again for $\lambda =0$:
\begin{equation}
    \left\langle \hat{f}(t)\hat{f}(t')\right\rangle^{(MF)} =0 \Rightarrow Q(t,t') =0
\end{equation}

Furthermore, we note that integrating out the field $\{f\}$ in  the joint distribution, Eq.~\eqref{eq:Pdef} we obtain
\begin{equation}
 \mathbb{P}\big(\{{x}\}_{0,t_f}\big)= \int D\hat{x} \exp\left(\hat{x}(t) \Big[(\partial_t +1)x(t)  - {\rm tanh}\big(\eta(t)  \big) \Big]+\, i\sigma^2 \hat{x}^2_k(t)/2\right)
\end{equation}
which corresponds to the path measure of the effective dynamics in Eqs.~\eqref{effective_dynamics}-\eqref{x_correlator}

One can now use the definition of the moment generating function to show that
\begin{eqnarray}\left\langle\dot{S}^{\textrm{res}}(t)\right\rangle&=&\left.\frac{1}{i}\frac{\delta}{\delta \lambda(t)}Z_S^{(MF)}[\lambda]\right|_{\lambda=0}\nonumber\\
&=&N\left.\left(\left\langle\dot{x}(t)\circ\big[\tanh\left(f(t)\right) - x(t)\big]\right\rangle^{(MF)}+\frac{J^2}{2}\frac{\delta}{\delta \lambda(t)}\int dt'\int dt''\,q(t,t')Q(t,t')\right)\right|_{\lambda=0}\nonumber\\&&-\left.\frac{NJ^2}{2}\left(\int dt'\int dt''\,\left(\frac{\delta}{\delta \lambda(t)}Q(t',t'')\left\langle x(t')x(t'')\right\rangle^{(MF)}+\frac{\delta}{\delta \lambda(t)}q(t,t')\left\langle \hat{f}(t)\hat{f}(t')\right\rangle^{(MF)}\right)\right)\right|_{\lambda=0}\nonumber\\
&=&N\left\langle\dot{x}(t)\circ\big[\tanh\left(f(t)\right) - x(t)\big]\right\rangle^{(MF)},\end{eqnarray}
where we used Eq.~\eqref{eq:qsollam0} and the fact that $\left\langle \hat{f}(t)\hat{f}(t')\right\rangle^{(MF)}$ to arrive at the last line. This completes the proof.

\section{Appendix B: Derivation of Eq.~(\ref{eq:dsc})\label{app:B}}
To compute the averaged entropy production rate we start from the definition and assume steady-state to find
\begin{equation}
\begin{aligned}   \langle \dot{s}^{(MF)} \rangle & \equiv \frac{1}{i}\frac{\partial}{\partial \lambda(t)} Z_S[\lambda]\Big|_{\lambda=0} \\ & =\frac{2}{\sigma^2}\left\langle \dot{x}(t)\times \Big[{\rm tanh}\big(\eta(t)\big) - x(t)\Big] \right\rangle \\&= \frac{2}{\sigma^2}\left.\frac{d}{d\tau}C_{ux}(\tau)\right|_{\tau=0}  \end{aligned}\label{eq:sig1}
\end{equation}
On the other hand, one can solve Eq.~\eqref{effective_dynamics} to find
\begin{equation}
    x(t)=\int^{\infty}_0d\tau e^{-\tau}\left(\xi(t-\tau)+u(t-\tau)\right)
\end{equation}
Due to the steady state assumption, one can use this to write
\begin{eqnarray}
    \frac{d^2}{dt^2}C_x(t)&=&\frac{d^2}{dt^2}\int^{\infty}_0d\tau\int^{\infty}_0d\tau'\,e^{-\tau-\tau'}\left(\left\langle\xi(-\tau)\xi(t-\tau')\right\rangle+\left\langle u(-\tau)u(t-\tau')\right\rangle\right)\nonumber\\
    &=&\frac{d}{dt}\int^{\infty}_0d\tau\int^{\infty}_0d\tau'\,e^{-\tau-\tau'}\frac{d}{d\tau}\left(\left\langle\xi(0)\xi(t-\tau'+\tau)\right\rangle+\left\langle u(0)u(t-\tau'+\tau)\right\rangle\right)\nonumber\\
    &=&\frac{d}{dt}\int^{\infty}_0d\tau'\,e^{-\tau'}\left(\left\langle\xi(0)\xi(t-\tau')\right\rangle+\left\langle u(0)u(t-\tau')\right\rangle\right)-\frac{d}{dt}C_x(t)\nonumber\\
    &=&\sigma^2\frac{d}{dt}e^{-t}+\frac{d}{dt}C_{ux}(t)-\frac{d}{dt}C_x(t)
\end{eqnarray}
where we used
\begin{equation}
    \frac{d}{dt}C_{ux}(t)=\frac{d}{dt}\int^{\infty}_0d\tau e^{-\tau}\left(\left\langle u(0)\xi(-(-(t-\tau)\right\rangle+\left\langle u(0)u(t-\tau)\right\rangle\right)\nonumber\\
\end{equation}
to arrive at the last line. We can now use Eq.~\eqref{initial_condition} to find
\begin{equation}
   \left. \frac{d}{dt}C_{ux}(t)\right|_{t=0}=\left.\frac{d^2}{dt^2}C_x(t)\right|_{t=0}-\frac{\sigma^2}{2}.
\end{equation}
Plugging this in into Eq.~\eqref{eq:sig1} imediately leads to Eq.~\eqref{eq:dsc}

\section{Appendix C: derivations of Eqs.~(\ref{EOM_correlation})-(\ref{initial_condition})}
Firstly, we note that by multiplying the DMFT equation by itself, one can get
\begin{equation}
    (\partial_t+1)x(t)(\partial_{t'}+1)x(t')=(u(t)+\xi(t))(u(t')+\xi(t'))
\end{equation}
Taking the average of this equation under steady-state and using time-translation invariance to write $\tau=t-t'$, one can rewrite this equation to
\begin{equation}
    -\partial^2_{\tau}C_x(\tau)+C_x(\tau)=\left\langle \tanh(\eta(\tau))\tanh(\eta(0))\right\rangle_{\eta}+\sigma^2\delta(\tau).
\end{equation}
One can then use Price's theorem to show that
\begin{equation}
  \frac{\partial}{\partial C_x(\tau)}\big\langle\Omega(\eta(\tau)) \Omega(\eta(0))\big\rangle_\eta = J^2\left\langle \tanh(\eta(\tau))\tanh(\eta(0))\right\rangle_{\eta}.
\end{equation}
Combining these two equations leads to Eqs.~\eqref{EOM_correlation}.

To derive the boundary conditions, we first note that Eq.~\eqref{EOM_correlation} corresponds to the Euler-Lagrange equation of the following Lagrangian for $t>0$:
\begin{equation}
    \mathcal{L}=\frac{\dot{C}_x(t)}{2}-V(C_x(t)|C_x(0)).
\end{equation}
One can than use methods from analytical mechanics to show that
\begin{equation}
    E=\frac{\dot{C}_x(t)^2}{2}+V(C_x(t)|C_x(0))
\end{equation}
is a constant of motion. We know on physical grounds that
\begin{equation}
    \lim_{t\rightarrow \infty}C_x(t)=\lim_{t\rightarrow \infty}\dot{C}_x(t)=0,
\end{equation}
and that $V(0|C_0)=0$. Combining these equations leads to
\begin{equation}
    \frac{\dot{C}_x(0^+)^2}{2}+V(C_x(t)|C_x(0))=0.\label{eq:eomdc}
\end{equation}

Secondly, we note that, due to time-translation symmetry, one has
\begin{equation}
    \dot{C}_x(0^+)=-\dot{C}_x(0^-). 
\end{equation}
This can only satisfy the first line of Eq.~\eqref{EOM_correlation} if
\begin{equation}
    \dot{C}_x(0^+)=-\frac{\sigma^2}{2}.
\end{equation}

Combining this with Eq.~\eqref{eq:eomdc} gives the initial conditions Eq.~\eqref{initial_condition}.

\section{Appendix D: Asymptotics of the entropy production}
In this section, we will derive the asymptotics of the entropy production, as given in Eq.~\eqref{eq:appr}. To do this, we will first show that
\begin{equation}\label{eq:Vappr}
    V(C_x(t)|C_x(0))\approx\begin{cases}
        \frac{(J^2-1)}{2}C_x(t)^2 &J\ll 1\\
        C_x(t)^4-\frac{(J-1)^2}{6}C_x(t)^2 & 0<J-J_c\ll 1\\
        \frac{8(\sqrt{C_0^2-C_x(t)^2}-C_0^2)+\pi C_x(t)-2\pi C_x(t)^2+2C_x(t)\left(3\tan^{-1}\left(\frac{\sqrt{C_x(t)^2-C_0^2}}{C_x(t)}\right)-\tan^{-1}\left(\frac{C_x(t)}{\sqrt{C_x(t)^2-C_0^2}}\right)\right)}{8\pi} & J\gg 1
    \end{cases}.
\end{equation}
One can then use the boundary condition, Eq.~\eqref{eq:dsV}, to show that
\begin{equation}   C_x(0)\approx\begin{cases}\frac{\sigma^2}{2\sqrt{1-J^2}} & J\ll 1\\          \left[\frac{  (J-1)^2}{2} + \frac{1}{2}\,\sqrt{(J-1)^4 -3\sigma^4 }\right]^{1/2} & 0<J-J_c\ll 1\\
         1-\frac{2}{\pi}+\sqrt{\left(1-\frac{2}{\pi}\right)^2+\frac{\sigma^4}{2}} &J-J_c\gg 1
    \end{cases}
\end{equation}
One can then plug this in into Eq.~\eqref{eq:dsV} to arrive at the final expressions, Eq.~\eqref{eq:appr}. The remainder of this section focusses on the derivation of Eq.~\eqref{eq:Vappr} in the three different regimes.

\subsection{$\mathbf{J\ll 1}$}
In the limit $J\ll 1$, the (co)variance associated with $\eta$ will generally become small and therefore the behavior of $\Omega(\eta(t))$ only becomes important around $0$, where one can do a lowest-order Taylor expansion,
\begin{equation}
    \Omega(\eta(t))=\frac{\eta(t)^2}{2}+O(\eta(t)^4)
\end{equation}
Therefore, we can write
\begin{eqnarray}
    \left\langle \Omega(\eta(0))\Omega(\eta(t))\right\rangle-\left\langle \Omega(\eta(0))\right\rangle\left\langle\Omega(\eta(t))\right\rangle&\approx& \frac{\left\langle \eta(0)^2\eta(t)^2\right\rangle-\left\langle \eta(0)^2\right\rangle^2}{4}\nonumber\\
    &=&\frac{J^4C_x(t)^2}{2},
\end{eqnarray}
where we used the fact that the distribution is Gaussian to arrive at the second line.
Plugging this into the definition of the potential, Eq.~\eqref{EOM_correlation} immediately leads to the first line of Eq.~\eqref{eq:Vappr}.

\subsection{$\mathbf{0<J-J_c\ll 1}$ and $\mathbf{\sigma\ll 1}$}
It can be shown that in the limit $\sigma= 0$ and $J\leq 1$ one has $C(t)=0$ for all $t$. Therefore, we can assume in this section that $C(t)$ is small and we can expand
\begin{equation}
    \Omega(\eta(t))\simeq \frac{\eta(t)^2}{2}-\frac{\eta(t)^4}{12}+\frac{\eta(t)^6}{45}+O(\eta(t)^8).
\end{equation}
This leads to
\begin{eqnarray}
\left\langle\Omega(\eta(0)\Omega(\eta(t))\right\rangle&\simeq&\frac{\left\langle \eta(0)^2\eta(t)^2\right\rangle}{4}-\frac{\left\langle \eta(0)^2\eta(t)^4\right\rangle}{12}+\frac{\left\langle \eta(0)^4\eta(t)^4\right\rangle}{144}+\frac{\left\langle \eta(0)^2\eta(t)^6\right\rangle}{45}\\
&=&\frac{J^6}{6}C(t)^4+\left(\frac{5}{2}J^6C_0^2-J^4C_0+\frac{J^2}{2}\right)C(t)^2,\label{eq:OmCor}
\end{eqnarray}
where we used the Gaussianity of the distribution. One can use this to show that
\begin{equation}
    V(C_0|C_0)\simeq \frac{8}{3}J^6C_0^4-J^4C_0^3+\frac{(J^2-1)}{2}C_0^2.
\end{equation}
In the limit of zero noise, this means that
\begin{equation}
    \frac{8}{3}J^6C_0^4-J^4C_0^3+\frac{(J^2-1)}{2}C_0^2\simeq0
\end{equation}
or
\begin{eqnarray}
    C_0&\simeq &\frac{3}{16J^3}\left(J-\sqrt{\frac{16-13J^2}{3}}\right)\nonumber\\
    &=&(J-1)-\frac{5}{6}(J-1)^2+O((J-1)^3),
\end{eqnarray}
where we did a Taylor expansion around $J=1$ in the second line. Plugging this back in into Eq.~\eqref{eq:OmCor} and doing a lowest order expansion leads to
\begin{equation}
    V(C|C_0)\simeq \frac{C^4}{6} - \frac{(J-1)^2}{6}\, C^2
\end{equation}

If we now use this equation of $V$ as an approximation for small but finite $\sigma$, we get the following boundary condtion:
\begin{equation}   \frac{C_0^4}{6} - \frac{(J-1)^2}{6}\, C_0^2 + \frac{\sigma^4}{8}=0
\end{equation}
Solving this equation for $C_0$, we obtain the physical solution
\begin{equation}
   C_0 = \left[\frac{  (J-1)^2}{2} + \frac{1}{2}\,\sqrt{(J-1)^4 -3\sigma^4 }\right]^{1/2}
\label{approximate_correlation_beyond_critical}
\end{equation}
 In the small noise regime, for $J-J_c<0.1$ the autocorrelation $C_0$ derived from our approximation agrees well with what can be obtained from the numerical solution of the self-consistency equation Eq. \eqref{initial_condition}. 
However, as long as $J-J_c$ gets larger, our approximation in Eq. \eqref{approximate_correlation_beyond_critical} overestimates the true value of $C_x(0)$, resulting in an overestimation of the entropy production rate.

\subsection{$\mathbf{J\gg 1}$}
In the limit $J\gg 1$ the variance of $\eta(t)$ diverges and therefore the averages of $\Omega(\eta(t))$ are dominated by their tales, where one can approximate
\begin{equation}
    \Omega(\eta(t))\approx \left|\eta(t)\right|-\ln 2.
\end{equation}
Therefore,
\begin{eqnarray}
    \left\langle \Omega(\eta(0))\Omega(\eta(t))\right\rangle-\left\langle \Omega(\eta(0))\right\rangle\left\langle\Omega(\eta(t))\right\rangle&\approx& {\left\langle \left|\eta(0)\right|\left|\eta(t)\right|\right\rangle-\left\langle \left|\eta(0)\right|\right\rangle^2}\nonumber\\
    &=&   \frac{8(\sqrt{C_0^2-C_x(t)^2}-C_0^2)+\pi C_x(t)}{8\pi J^2}\nonumber\\&&+\frac{C_x(t)\left(3\tan^{-1}\left(\frac{\sqrt{C_x(t)^2-C_0^2}}{C_x(t)}\right)-\tan^{-1}\left(\frac{C_x(t)}{\sqrt{C_x(t)^2-C_0^2}}\right)\right)}{4\pi J^2},
\end{eqnarray}
Plugging this into the definition of the potential, Eq.~\eqref{EOM_correlation} immediately leads to the last line of Eq.~\eqref{eq:Vappr}.

\end{document}